\documentclass[ ] 
{article}

\usepackage{epsf}

\begin{document}

\title{Evidence for the Strangeness S=+1 Pentaquark from 
LEPS and CLAS Experiments}

\author{K.H. Hicks\\
  {\it Department of Physics, Ohio University, Athens, OH 45701}
}

\date{ October 26, 2003 }

\maketitle

\begin{abstract}
There are now several experimental collaborations that have seen 
evidence for a narrow state in the mass spectrum of the ($nK^+$) 
system.  Two of these experiments, from the LEPS collaboration 
in Japan and the CLAS collaboration in the USA, are described 
briefly.  Both use similar photoproduction reactions with a 
$K^+K^-$ pair in the final state.  In addition, data from the 
CLAS collaboration for the $\gamma p \to K_s K^+ n$ reaction 
are presented for the first time, which has no prominant peak 
in the ($nK^+$) mass spectrum when the $K_s$ angle is 
limited to forward angles.
\end{abstract}


\section{Introduction}

Until recently, it was thought that the pentaquark, defined 
as a particle resonance with four quarks and one anti-quark, 
did not exist.  This belief was based on exhaustive searches 
for a strangeness $S=+1$ resonance in the 1970's \cite{PDG} 
yet only $S=-1$ particles were found.  This result was a 
bit surprising because the rules of QCD do not forbid the 
existence of pentaquarks.

Nonetheless, progress was made in theoretical studies of the 
soliton model of the nucleon \cite{praszalowicz} which 
predicted, in addition to the usual octet and decuplet, 
an anti-decuplet ($\bar{10}$) 
of baryons that includes a $S=+1$ particle.  Real progress 
was made when Diakonov, Petrov and Polyakov \cite{diakonov} 
predicted the mass of this particle using symmetries of the 
chiral soliton model and the key identification of the $S=0$ 
baryon of the $\bar{10}$ with the spin $\frac{1}{2}$, 
$P_{11}$ nucleon resonance at 1710 MeV.  
In this model, the mass of the 
pentaquark (called the $Z^+$ in Ref. \cite{diakonov} but 
since renamed the $\Theta^+$ by the authors) was predicted 
to have a specific mass of 1530 MeV and a narrow width 
($<15$ MeV). This motivated experimters to look again for 
this $S=+1$ particle in already-existing data.

\section{First Results}

\begin{figure}
\centerline{
  \epsfxsize=20pc
  \epsffile{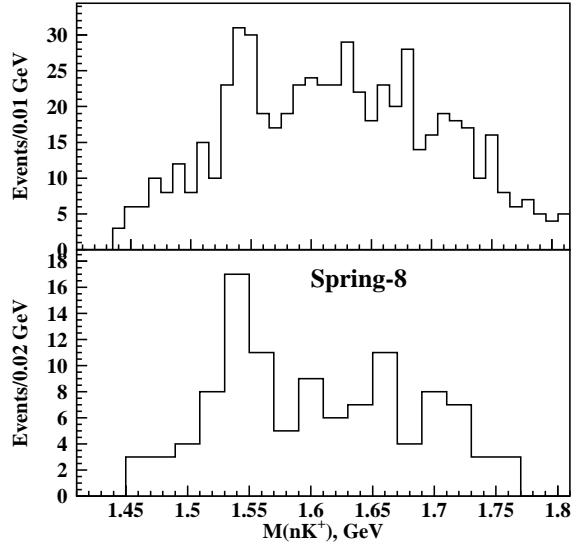}
}
  \caption{Comparison the invariant mass of the neutron-$K^+$ 
	system, $M(nK^+)$ for the CLAS data (top) and LEPS data 
        (bottom).}
  \label{fig:thplus}
\end{figure}

The first evidence for the $\Theta^+$ pentaquark was reported 
in October, 2002 \cite{nakano} by the LEPS collaboration in 
Japan.  The reaction is $\gamma n \to K^-\Theta^+ \to K^-K^+n$ 
where the neutron is bound inside a carbon nucleus, and only 
the $K^+$ and $K^-$ were detected at forward angles 
($\theta_{LAB}(K) < 30^\circ$).
The results are now published \cite{LEPS} and details 
of the measurement can be found there. The final spectrum 
is shown in Fig. \ref{fig:thplus} (bottom).

After the announcement by LEPS, the CLAS collaboration started 
to look for the $\Theta^+$ in existing photoproduction data on 
a deuterium target. The advantage of a deuterium target is 
that a kinematically complete reaction can be measured, in 
contrast to the inclusive kinematics of the carbon target. 
The reaction $\gamma d \to K^- \Theta^+ p \to K^- K^+ p (n)$ 
was analyzed, where the neutron was deduced by the missing 
mass technique.  Details of this result were first presented 
in February 2003 \cite{hicks}, followed by a more complete 
report in May \cite{stepanyan}.  Details 
can be found in Ref. \cite{CLAS}.  The final spectrum is 
shown in Fig. \ref{fig:thplus} (top).

Comparing both figures, the prominant feature is a peak at 
the same mass, 1.54 GeV, with a width of $<25$ MeV.  The 
width of the peak is consistent with the known resolution 
of each experiment.  This suggests that the intrinsic resolution 
of the resonance is smaller than the measured widths, although 
the small statistics prevents a definite conclusion.  The 
shape of the background under the peaks is described in the 
respective references (\cite{LEPS,CLAS}).  As long as the 
conservation laws of baryon number and strangeness hold, the 
resonance peak is a pentaquark made up of ($uudd\bar{s}$) 
construction.  This is consistent with the $\Theta^+$ 
prediction of Diakonov {\it et al.}, but until the spin and 
parity of this resonance is measured, we can not be sure 
that it is, indeed, the $\Theta^+$ particle as predicted.

\begin{figure}
\centerline{
  \epsfxsize=20pc
  \epsffile{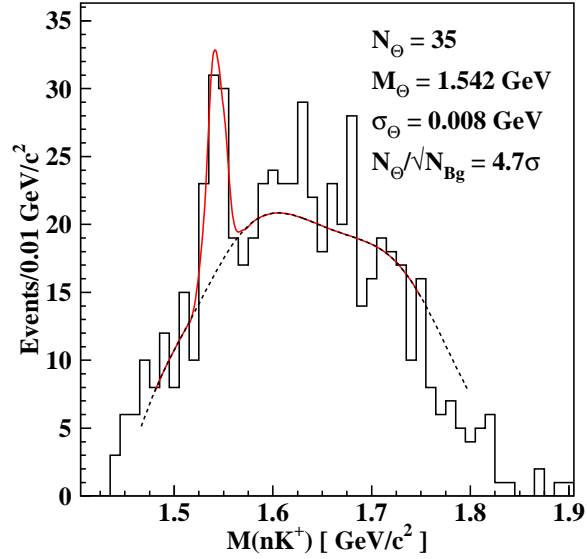}
}
  \caption{Peak fit using a MC background shape for the $M(nK^+)$ 
           spectrum.}
  \label{fig:CLASfit}
\end{figure}

The CLAS data are shown again in Fig. \ref{fig:CLASfit} along 
with a fit to the peak.  The background has been modeled by 
s-wave (non-resonant) photoproduction of $K^+K^-$ pairs.  The 
production cross section is just the phase space of 3-body 
($K^+K^-$ from a nucleon) and 4-body ($K^+K^-$ from both 
nucleons in deuterium).  Using this background shape, the 
fit gives a statistical significance of 4.7 $\sigma$, 
calculated as a fluctuation of the excess above the background 
shape (1 $\sigma$ = $\sqrt{N_{Bg}}$, where $N_{Bg}$ is the 
number of counts in the background within $\pm 20$ MeV of 
the peak's central mass).  The uncertainty in the statistical 
significance depends on the shape of the background, and 
different choices of background give values ranging from 
4.6 to 5.8 $\sigma$.

If the $\Theta^+$ exists, then it should be produced in a 
variety of reactions.  In addition to photoproduction from 
the neutron, which must be bound in a nucleus, it is natural 
to look for photoproduction from the proton.  Preliminary 
results on the reaction $\gamma p \to K^-\pi^+\Theta^+$ 
have already been presented \cite{stepanyan}.  Another 
reaction is $\gamma p \to \bar{K}^0 \Theta^+$ which was 
published by the SAPHIR collaboration \cite{SAPHIR} (which 
are also reported in these proceedings).

\section{New Experimental Results}

\begin{figure}
\centerline{
  \epsfxsize=20pc
  \epsffile{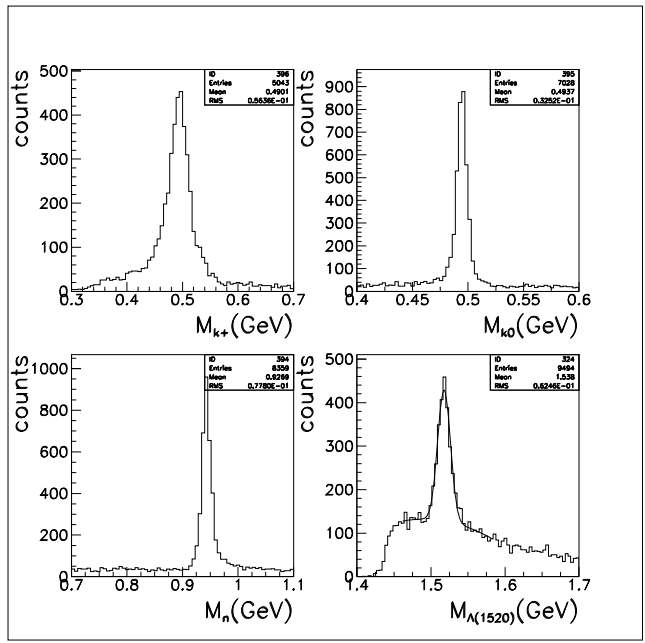}
}
  \caption{The masses calculated from coincident $K^+$, $\pi^+$ 
	and $\pi^-$ particles in photoproduction from a proton 
	target at CLAS.}
  \label{fig:masses}
\end{figure}

We present here {\it preliminary} results of the CLAS 
collaboration \cite{INFN} for analysis of the reaction 
$\gamma p \to \bar{K}^0 K^+ n$ 
where the neutron is measured by the missing mass technique.  
Fig. \ref{fig:masses} shows the the mass calculated from 
the momentum and velocity of detected particles.  The $K^+$ 
(top left) is detected directly.  The $K^0$ mass (top right) 
is from $K_s$ decay, made from the invariant mass of a detected 
$\pi^+\pi^-$ pair.  The $n$ mass (bottom left) is from the 
missing momentum and energy.  The $\Lambda (1520)$ mass 
(bottom right) is from the invariant mass of the deduced 
4-vectors of the $n$ and the $K_s$. All mass spectra show 
clear peaks with very 
little background, showing that particle identification is 
quite good.  Several thousand good ($K_sK^+$) events are 
identified for further analysis.

One must be careful in the analysis of the proton data because 
the $K_s$ is a linear combination of $K^0$ and anti-$K^0$.  
Hence, the strangeness of the reaction is not uniquely identified 
by the $K_s$ particle, and other reactions can produce the same 
final state.  As shown above, the $\gamma p \to K^+ \Lambda^*$ 
reaction, where $\Lambda^* \to K^0 n$, has the same final state 
as $\gamma p \to \bar{K}^0 \Theta^+$ where the $\Theta^+$ can 
decay to $K^+n$.  Also, since the $K_s$ is determined from a 
$\pi^+\pi^-$ pair, the reaction $\gamma p \to K^+ \Lambda^*$ 
followed by $\Lambda^* \to \Sigma \pi \to \pi^+\pi^- n$ has the 
same particles in the final state.  

\begin{figure}
\centerline{
  \epsfxsize=20pc
  \epsffile{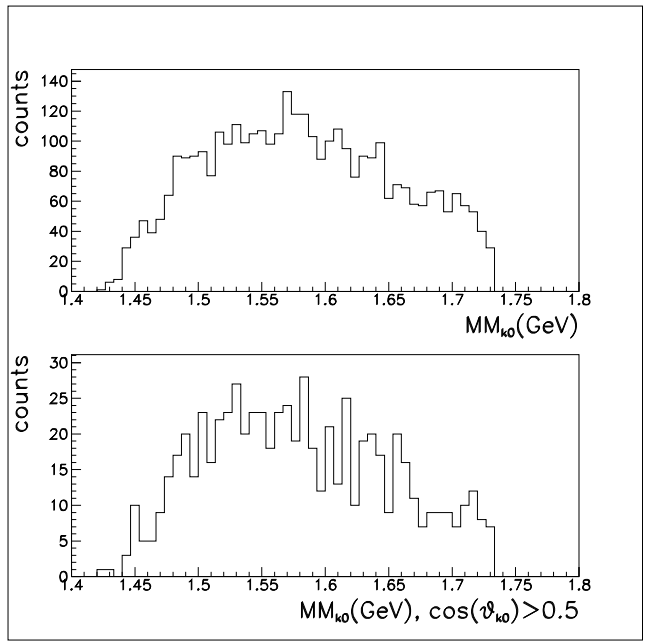}
}
  \caption{The missing mass spectrum for $K^0_s$ production 
	from the reaction $\gamma p \to \bar{K}^0 n$ and 
	event selection as described in the text\cite{INFN}.}
  \label{fig:final}
\end{figure}

After rejecting events where a $\pi n$ invariant mass 
equals the $\Sigma$ mass, and also events in the 
$\Lambda (1520)$ peak (see Fig. \ref{fig:masses}), 
the missing mass of the $K_s^0$ spectrum is shown in 
Fig. \ref{fig:final}.  This spectrum should show a peak at 
the mass of the $\Theta^+$ if this state is produced with 
a cross section sufficiently larger than the non-resonant 
background.  If the $\Theta^+$ is produced in a $t$-channel 
process, then selecting events with forward angles of the 
$K_s$ ($\cos \theta_{K_s} > 0.5$ where the angle is in 
the photon-proton center of mass system) could enhance the 
signal over the background.  The bottom plot in Fig. 
\ref{fig:final} shows the spectrum after this event cut.
In both cases, no prominant peak is observed.

\section{Summary and Conclusions}

The lack of a strong signal in the {\it preliminary} analysis 
of the $\gamma p \to K_s^0 K^+ n$ reaction at CLAS is 
surprising, considering that there are several measurements 
that have shown strong evidence for the $\Theta^+$.   
One possible explanation is that 
the coupling constant at the $K^*p\Theta^+$ vertex is small, 
giving a small cross section in the $t$-channel.  (The $K^*$ 
vector meson is necessary as a virtual particle at forward 
angles since the neutral kaon couples to the photon primarily 
through a magnetic M1 transition).  Of course, another 
possibility is that the $\Theta^+$ does not exist, and that 
the other experiments are just unfortunate statistical 
fluctuations.  (At the time of writing this paper, the latter 
possibility seems less likely, as more reports supporting the 
$\Theta^+$ have been announced.)

A full understanding of the experimental situation for 
evidence of the $\Theta^+$ must await future measurements. 
It is difficult to be patient at a time where excitement 
surrounds the announcement of a new particle, which could 
be the beginning of a new class of particles (pentaquarks). 
However, we must be cautious, and let the facts emerge.  
If the $\Theta^+$ exists, then experiments with better 
statistics and more understanding of the background will 
find clear evidence for this particle.  Until then, there 
are positive signs but no definite conclusions regarding 
the existence of pentaquarks.


\subsection*{Acknowledgments}
This work would not be possible without the hard work of many 
people in the LEPS and CLAS collaborations. The financial support 
of the U.S. Department of Energy (DOE), the National Science Foundation 
(NSF) and the Japanese Ministry of Education, Culture, Science and 
Technology (MEXT) is gratefuly acknowledged.  The analysis in Figs. 3-4 
was supported by the Italian Instituto Nazionale de Fisica Nucleare.

\end{document}